\documentclass[11pt,twoside]{article}
\usepackage{asp2010}

\resetcounters

\begin{document}

\title{Development of a VO Registry Subject Ontology using Automated Methods}
\author{Brian~Thomas \affil{National Optical Astronomy Observatory, Tucson, AZ 85719}}

\begin{abstract}
We report on our initial work to automate the generation of a domain ontology using subject fields of resources held in the Virtual Observatory registry. Preliminary results are comparable to more generalized ontology learning software currently in use. We expect to be able to refine our solution to improve both the depth and breadth of the generated ontology.
\end{abstract}

\section{Introduction}

Ontologies promise a rich user interaction with large amounts of data. They may be used to map the heterogeneous semantics which various data repositories use to label their data into a common ontology (or set of ontologies) which describe the aggregate of all available data. This common ontology may in 
turn then be used to create complex queries which can precisely describe the data of interest using concepts which are familiar to the end user scientist.
 
The development of such an ontology is a non-trivial matter however. Problems include the amount of human effort required to both populate and keep up to date individuals (instances) of the ontology (more data may be added after the initial ontology is developed). Furthermore, there are maintenance costs associated with maintaining the ontology itself. The semantics in use at the various data repositories will evolve (ex. new classes of subjects are added) and the common ontology must evolve to encompass these changes.

The Virtual Observatory (VO) registry presents an interesting test case for developing automated methods to do these tasks. The VO registry contains approximately 30,000 resources, which are simply elements of the VO, such as organizations, data collections or services, that can be described in terms of who curates or maintains it and which can be given a name and a unique identifier. Each resource is allowed to be labeled with one or more subject fields. While its entries each conform to a prescribed data model, the semantics of the VO registry data model \citep{Han_2007} which describe the content (subject) of the data are not constrained and each publisher is free to label the subject of the data as they wish. 

Our motivation is to create a subject ontology for VO resources which will expand the applicable search results beyond a simple matching against existing terms taking into account synonyms and hypernyms (parent concepts). For example, a search for resources with the subject of ``star" should also turn up all resources not explicitly labeled such as resources which have sub-classed star subjects like ``early-type stars" or ``Wolf-Rayet". To enable such a query, we plan to map existing resources into an subject ontology. 

\section{Methodology}

There are many approaches to generation of ontologies in the literature.
Effort is generally directed towards developing generalized solutions, which may handle  generation of an ontology from any selected corpus of text regardless of the domain(s) to which they may belong. In our case, we are harvesting information about VO registry resources via their subject fields, and this supplies some advantages not enjoyed by others. First, the subject text typically holds only one or more mostly noun keywords rather than whole sentences or paragraphs ($\sim$5\% of all subject fields contain sentences). We may further assume that all subject text belongs to the same domain. In other words, it is reasonable to assume, for example, that the term ``star" is always considered to be an astrophysical object rather than meaning ``an asterisk". This assumption, coupled with ignoring the small amount of subject text which are more than keywords, allows us to side-step the use of fancier methods which are used to extract concepts. 

\citet{Lon_2010} have outlined a general methodology for ontology generation which we have adapted here. They describe a series of primary steps which involve first the selection of concepts  and then the retrieval of relationships from a corpus of base text documents and a single source ontology (we don't pursue their last step of constraint discovery here). 

We have chosen to use the IVOA Thesaurus, ``IVOAT" \citep{Hes_2008}, as the basis for our source ontology. This is an ideal choice as it covers a broad range of concepts in Astronomy similar to the range of subjects in the VO registry. The IVOAT is serialized as a SKOS vocabulary, so to produce the ontology we have used an XSLT stylesheet to transform it, using a simple mapping of transforming SKOS concepts directly into OWL classes, importing SKOS broader relationships to create an ``is-a" hierarchy and the {\tt prefLabel} and {\tt altLabel} elements for each concept to record any known synonyms.

In this work, concept selection involves the harvesting of subject text into a corpus of unique instances after filtering out sentence text. This produces a list of about 1100 text instances. We next utilize a simple tokenizer to extract subject concepts from the corpus. Tokenization parses out concepts from text using a small set of regular expressions we have developed. These expressions serve to parse concepts from comma, semi-colon or space delimited lists, can change casing of concepts from plural, reformat concept text into standard English from specialized formatting (``star:binary" to ``binary\_star" for example). We then filter this list to drop any unusual acronyms and/or contractions (such as ``cdfsagncxo"). We have adapted the filtering methodology of \citet{Yan_2008}, and our filtering is done by first referencing a small local domain dictionary (of Astronomical terms) followed by queries to Wikipedia and then WordNet (we differ from Yang \& Callan in that they utilized Google search instead). Any word we fail to identify in any one or more of these sources, is filtered out of candidate subject concepts list. Filtering in this manner results in a list of $\approx$450 concepts.

We assemble the list of filtered concepts into an initial, ``flat" ontology (ie. all concepts become classes which inherit from {\tt owl:Thing}) and proceed to merge this ontology with our source ontology,  by either making direct lexical matches between named classes (or their synonyms), or by indirect matching using hypernyms from the domain dictionary to identify any possible superclasses in the ontology we might use.

\section{Results}

\begin{figure}[t]
\epsscale{0.70}
\plotone{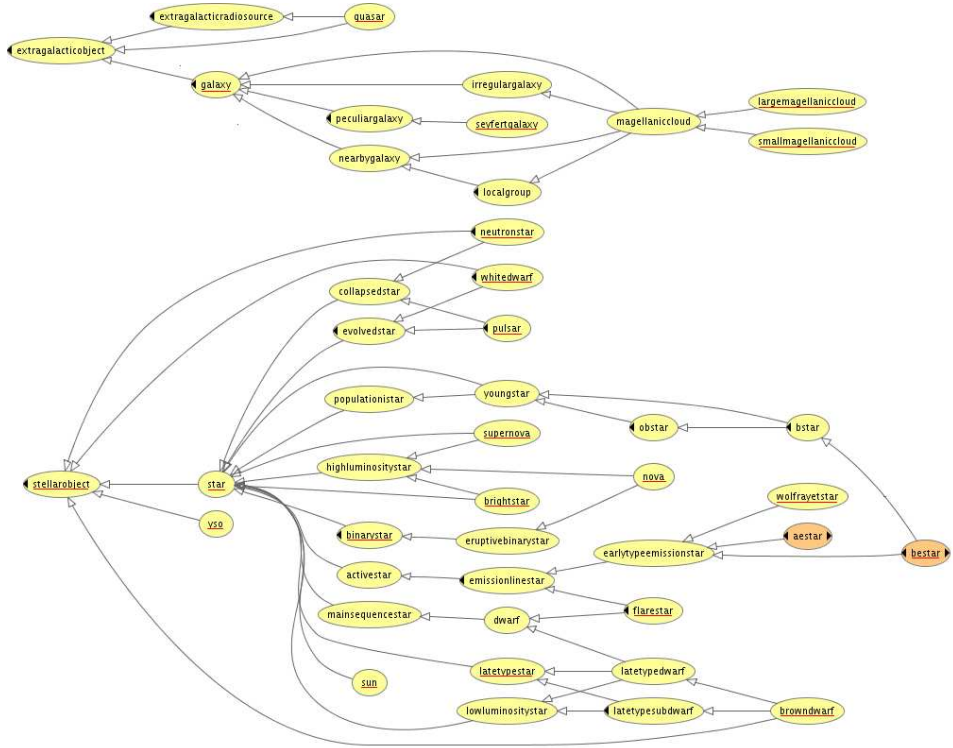}
\caption{Extragalactic and stellar object portions of the generated subject ontology. Red underlined items indicate concepts contained within the subject corpus, other subjects are pulled from the IVOAT source ontology during the merge stage.} \label{P074-fig-1}
\end{figure}

Figure~\ref{P074-fig-1} shows a portion of our subject ontology generated, with the classes underlined in red representing those classes which have made direct matches to subject concepts. Classes lacking underlining are simply imported from the source IVOAT ontology. This figure shows that for stellar and extragalactic concepts we have achieved some success, but this diagram only represents a small fraction of the corpus of subject terms ($\sim$3\%). How do we measure our work more meaningfully?

There are a number of more quantitative measures which one might use to gauge the performance of ontology learning software (ie. software which auto-generates ontologies). \citet{Zou_2009} give a review of many current measures. Because this is a preliminary work, we have chosen to simply apply a structural evaluation, similar to the class match measure of \citet{Ala_2006}, which evaluates the coverage of an ontology of the ``sought terms".

To obtain a scoring result for this measure, we evaluated all subject concepts which were successfully merged into source ontology. For each direct match between a class in the ontology and a subject concept we give a value of 1 (a direct match) and a value of 0.5 for an indirect match which occurs when we match a subject class by using a parent concept of a subject. The overall score is a ratio of the sum of these values divided by the number of subject concepts we made available for ontology generation. This score may thus range from 0 (no matches whatsoever) to 1 (all concepts directly matched). Using this measure we obtained a score of 0.32, which is low. Current ontology learning software averages about 0.3-0.5 (and sometimes even better) when generating an ontology from a corpus of 700 to 1000 sentences \citep[see][]{Zou_2009}.

Where might problems lie in our approach? A deeper look at the body of subject concepts shows that there are still some failures at parsing the grammar in the subject corpus, and sometimes we have split the text too far as for ``solar system" which becomes the separate concepts "solar" and "system". Other problems which lower the score include failed dictionary lookups for synonyms or hypernyms (such as for ``globular\_cluster" vs IVOAT class ``globularStarCluster") as well as having subject concepts which do not exist in the source ontology, nor have any matching hypernyms and therefore cannot be merged in. 

Nevertheless, by this scoring measure, this software has a performance comparable to the lower end of the current average ontology learning software available.

\section{Summary}

We have shown that some reasonable progress may be made towards the automated generation of a subject ontology for the VO registry. Results show that we have comparable ballpark performance to more generalized solutions which construct ontologies from text corpi. 

Because we are operating in a single domain, which contains many specialized concepts, we should be able to outperform these solutions. Possible directions to help increase the depth and breadth of the subject ontology include using WordNet to enhance the dictionary lookups of synonyms and hypernyms during the concept selection and merging stages, improving our local Astronomical dictionary to include more technical terms and hypernyms and using additional source ontologies, such as Ontology of Astronomical Object Types \citep{Der_2009}.


\bibliography{P074}

\begin{thebibliography}{}
\expandafter\ifx\csname natexlab\endcsname\relax\def\natexlab#1{#1}\fi
\expandafter\ifx\csname url\endcsname\relax
  \def\url#1{\texttt{#1}}\fi
\expandafter\ifx\csname urlprefix\endcsname\relax\def\urlprefix{URL }\fi
\providecommand{\eprint}[2][]{\url{#2}}

\bibitem[{Alani \& Brewster(2006)}]{Ala_2006}
Alani, H., \& Brewster, C. 2006, in Proceeding of 4th International EON
  Workshop, 15th International World Wide Web Conf. (Edinburgh: WWW)

\bibitem[{Derriere(2009)}]{Der_2009}
Derriere, S. e.~a. 2009, Ontology of astronomical object types.
  \urlprefix\url{http://www.ivoa.net/internal/IVOA/IvoaSemantics/OWLDOC-Object%
Types_tar.gz}

\bibitem[{Hanish(2007)}]{Han_2007}
Hanish, R. 2007, Resource metadata for the virtual observatory.
  \urlprefix\url{http://www.ivoa.net/Documents/latest/RM.html}

\bibitem[{Hessman(2008)}]{Hes_2008}
Hessman, F. 2008, A modified version of the 1993 iau thesarus.
  \urlprefix\url{http://volute.googlecode.com/files/ivoa-vocab-1.0.tar.gz}

\bibitem[{Lonsdale(2010)}]{Lon_2010}
Lonsdale, D. e.~a. 2010, Data and Knowledge Engineering, 69, 318

\bibitem[{Yang \& Callan(2008)}]{Yan_2008}
Yang, H., \& Callan, J. 2008, in Proceedings of the 2008 international
  conference on Digital government research, DG.O

\bibitem[{Zouaq \& Nkambou(2009)}]{Zou_2009}
Zouaq, A., \& Nkambou, R. 2009, Trans. Knowl. Data Eng., 21, 1559

\end{thebibliography}

\end{document}